\documentclass[letterpaper,pra,twocolumn,showpacs,superscriptaddress]{revtex4}

\usepackage{graphicx}
\usepackage{amsmath}
\usepackage{amssymb}

\begin{document}

\title{Generation of Polarization Squeezing with Periodically Poled KTP at 1064~nm}

\author{M.Lassen$^{1}$, M.Sabuncu$^{1,2}$, P.Buchhave$^{1}$ and U.L.Andersen$^{1,2}$}

\address{$^{1}$ Department of Physics, Technical University of Denmark,
Building 309, 2800 Lyngby, Denmark\\
$^{2}$Institut f\"{u}r Optik, Information und Photonik, Max-Planck
Forschungsgruppe, Universit\"{a}t Erlangen-N\"{u}rnberg,
G\"{u}nther-Scharowsky str. 1, 91058, Erlangen, Germany}

\email{mlassen@fysik.dtu.dk} 

\date{\today}

\begin{abstract}
We report the experimental demonstration of directly produced
polarization squeezing at 1064 nm from a type I optical parametric
amplifier (OPA) based on a periodically poled KTP crystal (PPKTP).
The orthogonal polarization modes of the polarization squeezed state
are both defined by the OPA cavity mode, and the birefringence
induced by the PPKTP crystal is compensated for by a second, but
inactive, PPKTP crystal. Stokes parameter squeezing of 3.6 dB and
anti squeezing of 9.4 dB is observed.
\end{abstract}


\maketitle

\section{Introduction}
The quantum properties of polarization states of light has recently
received a great deal of attention. It is not only interesting from
a fundamental point of view, but it also has some practical
relevance since it facilitates the execution of various quantum
information protocols: A future quantum information network will
probably consist of nodes of atoms, where the quantum information
are processed, linked by optical channels~\cite{Cirac1995}. The
transfer of quantum information from atoms to photons and photons to
atoms is made possible using the quantum properties of the different
polarization states \cite{Hald1999}.

Several methods for generating polarization squeezed states have
been proposed and experimentally realized. For example, making use
of the nonlinearity provided by Kerr-like media, such as in
nonlinear fibers and atomic media, or by combining a dim quadrature
squeezed beam generated by an optical parametric amplifier (OPA)
with a bright coherent beam on a polarizing beam splitter
(PBS)~\cite{Grangier1987,Hald1999,Korolkova2002,Heersink2003,Heersink2005,Bowen2002,
Josse2003,Sherson2006,Andersen2003}.

In the works by Heersink et al.~\cite{Heersink2003,Heersink2005},
polarisation squeezed light was generated by combining two
orthogonally polarisation components inside a spatial mode supported
by a fiber. It means that the polarisation squeezed state was
directly generated in the nonlinear medium. In contrast, in previous
approaches where optical parametric amplification has been used, the
polarisation squeezing was produced by combining a coherent beam
with a quadrature squeezed beam on a polarizing beam splitter. This
method is limited by the losses of the beam splitter and imperfect
spatial mode overlap in the beam splitter. In this paper we overcome
this problem by injecting the coherent beam into the cavity along
with the squeezed beam but in an orthognal polarization mode. Since
both beams are supported by the same spatial cavity mode the spatial
overlap between them is perfect, and thus the polarisation squeezing
is optimized. Also in this paper we demonstrate for the first time
squeezing at 1064~nm using periodically poled KTP. This crystal has
proven to be superior for squeezed state generation due to the
absence of nonlinear absorption effects. It has been use to squeeze
light at 532~nm \cite{Andersen2002},
795~nm~\cite{Tanimura2006,Hetet2007}, 860~nm~\cite{Takeno2007} and
946~nm~\cite{Aoki2006} but so far not at 1064~nm.

\section{Polarization squeezing}

The polarization state of light can be described by the four Stokes
operators $\hat{S}_{0}$, $\hat{S}_{1}$, $\hat{S}_{2}$ and
$\hat{S}_{3}$, where $\hat{S}_{0}$ represents the beam intensity
whereas $\hat{S}_{1}$, $\hat{S}_{2}$ and $\hat{S}_{3}$ characterize
its polarization and form a cartesian coordinate system . If the
Stokes vector points in the direction of $\hat{S}_{1}$,
$\hat{S}_{2}$ or $\hat{S}_{3}$ the polarized part of the beam is
horizontally, linearly at 45$^\circ$, or right-circularly polarized,
respectively \cite{Stokes,Siegman}. It is well known that the
polarization of a light beam is a property which is limited by
quantum noise and that it is possible to achieve polarization states
below the standard quantum noise limit (QNL). This was first
suggested in the work by Cirkin \emph{et al.} in 1993
\cite{Chirkin2003}, where the Heisenberg inequalities for the Stokes
operators were derived by making use of their cyclical commutation
relations:
\begin{equation}
V_{\hat{S}_i} V_{\hat{S}_j}\geq \varepsilon_{ijk}|\langle
\hat{S}_k^2\rangle|,
\end{equation}
where $i,j,k=1,2,3$. This means that intrinsic quantum fluctuations
of the different Stokes operators exist. The variances of the Stokes
operators, $V_i=\langle \hat{S}_i^2\rangle-\langle
\hat{S}_i\rangle^2$, can be expressed in terms of the amplitude and
phase quadrature operator variances, $V^\pm_{s,p}$, where $\pm$
refers to squeezing and anti-squeezing of the s- and p-polarized
states, respectively. Assuming that the s- and p-polarized states
are uncorrelated it can be shown that the variance of the different
Stokes operators are given by \cite{Schnabel2003}:
\begin{eqnarray}\label{SHGmodes}
V_0&=&V_1=\alpha_s^2 V^+_s + \alpha_p^2 V^+_p\nonumber\\
V_2(\theta)&=& \cos^2(\theta)\big(\alpha_s^2V^+_p +
\alpha_p^2V^+_s\big)\nonumber\\
&&+\sin^2(\theta)\big(\alpha_s^2V^-_p
+ \alpha_p^2V^-_s\big)\nonumber\\
V_3(\theta)&=&V_2(\theta-\pi/2),
\end{eqnarray}
where $\theta$ is the phase between the s-polarized and p-polarized
beams and $\alpha_{s}$, $\alpha_{p}$ are the classical amplitudes.
In our experiment we use a phase difference of $\theta=0$.

In the case of an s-polarized bright amplitude squeezed beam and
vacuum in the orthogonal polarization mode we expect the variances
to be: $V_0=V_1<1$ and $V_2=V_3=1$, where we have normalized the QNL
and amplitudes. Note that this is not polarization squeezing, but
simply amplitude squeezing. The generation of polarization squeezing
requires the presence of vacuum squeezing along with a bright beam.
For example, considering the case of a dim s-polarized squeezed beam
and a strong p-polarized coherent beam the expected variances will
be: $V_0=V_1=1$, $V_2<1$ and $V_3>1$ (see Eq.~\ref{SHGmodes}), and
thus the state is polarization squeezed.

The polarization states can also be visualized in a 3-dimensional
diagram, the so-called Poincar\'{e} sphere. 
The mapping of the state onto the Poincar\'{e} sphere provides the
full characterization. A recent example of such a mapping of the
polarization state was demonstrated by Marquardt \emph{et al.}
\cite{Marquardt2007}. The polarization plane on which the classical
stokes parameter is perpendicular defines the polarisation squeezing
of the state in case of large coherent
excitation~\cite{Heersink2005}. Therefore, if the light is
classically  S$_1$ polarized, the "dark" polarization plane is
spanned by the S$_2$ and S$_3$ parameters, and the squeezing and
anti-squeezing of the quantum polarization is to be found in this
plane. Using this simple observation, the analogy between
polarization squeezing and quadrature squeezing is obvious
\cite{Walls1994,Bachor2004}.

\section{Generation and detection of polarization squeezing}
\begin{figure}[tbp]
\begin{center}
\includegraphics[width=0.4\textwidth]{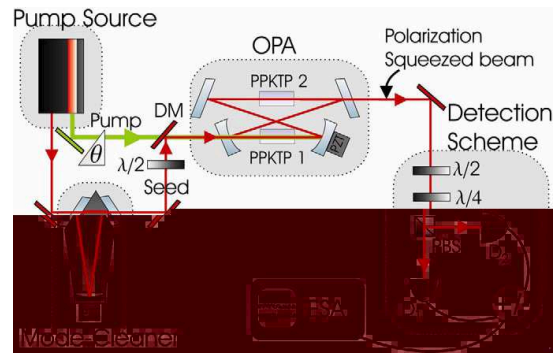} \caption{Schematics of
the experimental setup to generate amplitude and polarization
squeezing. PBS: polarizing beam splitter. DM: dichroic mirror.
$\lambda/2$: half-wave plate. $\lambda/4$: quarter-wave plate. Pol.
SQL: polarization squeezed light. $\theta$: phase between pump and
seed. The PPKTP1 crystal is the squeezing crystal, while PPKTP2 is
the birefringence compensating crystal.} \label{OPA-Scheme}
\end{center}
\end{figure}

Our squeezing setup is depicted in Fig.~\ref{OPA-Scheme} and
consists of an empty cavity, an optical parametric amplifier and a
detection scheme. The laser source, a cw solid-state monolithic YAG
laser, Diabolo from Innolight, provides 800~mW at 532~nm and 450~mW
power at 1064~nm. The 1064~nm beam from the laser is directed
through an empty ring cavity, a so-called mode-cleaner (MC), which
filters out the intensity and frequency noise of the laser above the
bandwidth of the MC. In addition the MC defines a high quality
spatial TEM$_{00}$ mode. A bandwidth of 2.7 MHz is measured and a
transmission greater than 90\% is obtained for the TEM$_{00}$ mode.

We used a bow-tie shaped cavity for the OPA in order to avoid
possible destructive interference that may result from a double
passed linear cavity.  Furthermore the circulating beam encounters
the passive loss in the crystal only once per round trip. The
nonlinear crystal used is a 1x2x10 mm$^3$  type I periodically poled
KTP (PPKTP) crystal manufactured by Raicol Inc. Our bow-tie cavity
consists of two curved mirrors of 25 mm radius of curvature and two
plane mirrors. Three of the mirrors are highly reflective at
1064~nm, $R>99.9$\%, while the output coupler has a transmission of
$T=15\%$. The transmittance of the mirrors at the pump wavelength,
532~nm, is more than 95$\%$. The crystal is placed in the smallest
beam waist, which is located between the two curved mirrors. In
order to maximize the conversion efficiency, corresponding to an
optimization of the Boyd-Kleinmann factor \cite{BK}, a beam waist of
20 $\mu$m is chosen. This was enabled by using a cavity length of
about 145 mm and setting the distance between the two curved mirrors
to be 28 mm. The OPA has a measured finesse of approximately 20, a
free spectral range (FSR) of 2 GHz and a cavity bandwidth of 100
MHz. From the finesse of the cavity we deduced an overall
intra-cavity loss of $L=1.0\pm0.2\%$.

We use the half-wave plate in front of the OPA to inject a beam that
has components of both s- and p-polarization, where the s-polarized
beam is a seed for the squeezed beam, and the p-polarized beam is
the coherent beam. Due to the birefringence of the nonlinear crystal
the s- and p-polarization is not simultaneously resonant. A
difference between the s- and p-polarization resonances is measured
to 0.5 FSR at a crystal temperature of 32$^\circ$C, which is the
optimal phase-matching temperature. By changing the temperature of
the crystal to 24$^\circ$C a simultaneous resonance of the s- and
p-polarization could be enabled. However, this temperature lies
outside the phase-matching bandwidth of the nonlinear crystal, which
was measured to be approximately 5$^\circ$C. We therefore use a
second, but identical, PPKTP crystal for compensating the
birefringence. We tune the temperature of the second crystal so that
the s- and p-polarization are simultaneously resonant. The cavity
was locked to resonance using the Pound Drever Hall locking
technique~\cite{PDH}.

Depending on the relative phase between the pump and the seed, the
seed is either amplified or de-amplified. We measure a maximum
amplification of 14 and a de-amplification of 0.38. The relative
phase is locked to de-amplification in order to generate an
amplitude quadrature squeezed beam. This beam is then directed to
the polarization measurement scheme \cite{Agarwal2003}.

In order to perform the measurement of the Stokes operators the
state propagates through a sequence of wave-plates and is
subsequently projected on a PBS. The intensities of the two outputs
are then measured and by subtracting the resulting photocurrents any
Stokes parameter can be accessed. If the waveplates are aligned so
that the polarization state is unchanged $\hat{S}_1$ is accessed. To
measure $\hat{S}_2$ the polarization of the beam was rotated by
45$^\circ$ with a half-wave plate, and if, in addition to the half
wave plate rotation, a quarter wave plate introduces a $\pi/2$ phase
shift between s- and p-polarization, $\hat{S}_3$ is measured. The
spectral densities of the resulting difference currents are measured
with an electronic spectrum analyser (ESA).

\begin{figure}[tbp]
\begin{center}
\includegraphics[width=0.45\textwidth]{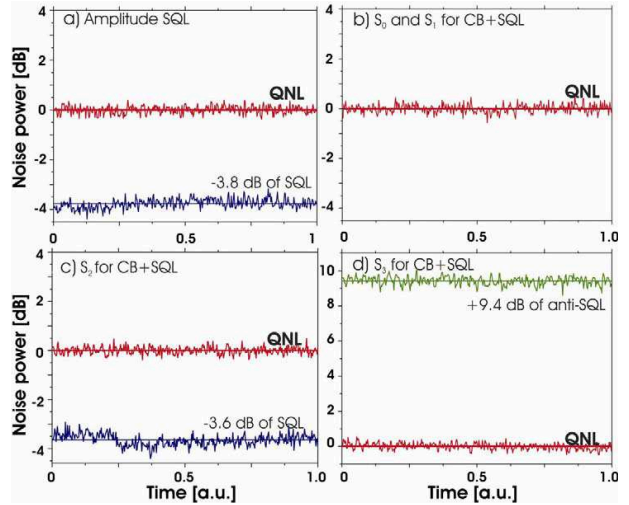}
\caption{a) Bright amplitude squeezed. b) $\widehat{S}_{0}$ and
$\widehat{S}_{1}$. c) $\widehat{S}_{2}$. d) $\widehat{S}_{3}$. The
measurements are taken at 14~MHz in zero span mode with RBW=300 kHz
and VBW=300 Hz. CB: coherent beam. SQL: amplitude squeezed light.}.
\label{POL SQL}
\end{center}
\end{figure}

The measurements were performed at a detection frequency of 14 MHz.
The detectors used are designed to be resonant at 14 MHz having a 1
MHz bandwidth. The variances have all been corrected for dark-noise
of the detectors, which is more than $12\pm0.2$~dB below the QNL,
and thus had a negligible effect on the noise spectrum. Each trace
depicted in Fig.~\ref{POL SQL} is normalised to the QNL and measured
with a resolution bandwidth (RBW) of the ESA set to 300 kHz and with
a video bandwidth (VBW) of 300 Hz. The calibration of the QNL is
done by operating the OPA without the pump. We then adjusted the
seed power to be equal to that of the squeezed beam when the pump is
turned on.

The total detection efficiency of our experiment is given by:
$\eta_{total}=\eta_{cav}\eta_{prop}\eta_{det}\eta_{OL}$, where
$\eta_{prop}=0.97\pm0.02$ is the propagation efficiency from the
cavity to the detectors, $\eta_{det}=0.95\pm0.02$ is the quantum
efficiency of the photo-detectors, $\eta_{cav}=T/(T+L)=0.94$ is the
quantum escape efficiency and $\eta_{OL}$ is the overlap efficiency.
Since both the seed and coherent beam are injected into the cavity
the efficiency is $\eta_{OL}\approx1$. The total detection
efficiency of our system is therefore $\eta_{total}=0.87\pm0.02$.

We first measured the polarization state of a single bright
amplitude squeezed beam without a coherent beam in the orthogonal
polarization mode. This is achieved by injecting all the OPA seed
light into the squeezed p-polarization mode and subsequently
measuring the amplitude of output with a single detector. The
variance of the measurement outcomes is displayed in Fig.~\ref{POL
SQL}a, and we see that amplitude squeezing of -3.8$\pm0.2$~dB
relative to the quantum noise limit is observed. Taking into account
the detection efficiency we infer an amplitude squeezing of
-5.0$\pm0.4$~dB.

Next we generated and measured the polarization state of a dim
amplitude squeezed beam combined with a strong coherent beam in the
orthogonal polarization mode. The seed light was then mainly
launched into the inactive polarization mode of the parametric
cavity and only a small amount into the squeezed mode. The measured
spectral densities associated with the three Stokes parameters are
shown in Fig.~\ref{POL SQL}b,c and d. Since the p-polarized coherent
beam was much more intense than the s-polarized squeezed beam, the
$\hat{S}_0$ and $\hat{S}_1$ parameters are at the QNL. The
$\hat{S}_2$ parameter is squeezed by -3.6$\pm0.2$~dB relative to the
QNL, while the $\hat{S}_3$ operator is anti-squeezed +9.4$\pm0.2$~dB
relative to the QNL. Taking into account the detection efficiency,
we infer squeezing and anti-squeezing values of -4.6$\pm0.4$~dB and
+10.0$\pm0.4$, respectively.

\section{Summary}
We have demonstrated the generation of polarization squeezed light
using periodically poled KTP. We generated -3.6$\pm0.2$~dB squeezing
in $\hat{S}_2$ and +9.4$\pm0.2$~dB anti-squeezing in $\hat{S}_3$
using a dim amplitude squeezed beam combined with a coherent beam.
To ensure an efficient spatial mode overlap between the coherent
state and the squeezed state, both modes were defined in the same
cavity in orthogonal polarization modes. We also produced bright
amplitude squeezing and measured -3.8$\pm0.2$~dB squeezing. To our
knowledge this is the first demonstration of squeezing at 1064~nm
using PPKTP.

This work was supported by the Danish Technical Research Council
(STVF Project No. 26-03-0304) and the EU project COVAQIAL (Project
No. FP6-511004).

\bibliographystyle{unsrt}

\begin{thebibliography}{99}

\bibitem{Cirac1995} J. I. Cirac and P. Zoller, Quantum Computations with
Cold Trapped Ions,  Phys. Rev. Lett. 74, 4091 (1995).
\bibitem{Hald1999} J. Hald, J. L. Sørensen, C. Schori and E. S. Polzik,
Spin Squeezed Atoms: A Macroscopic Entangled Ensemble Created by
Light, Phys. Rev. Lett. 83 1319 (1999).
\bibitem{Stokes} G. G. Stokes, On the composition and resolution of
streams of polarized light from different sources, Trans. Camb.
Phil. Soc. 9, 399 (1852).
\bibitem{Siegman} A. E. Siegman, LASERS, University Science Books, (1986)
\bibitem{Chirkin2003} A. S. Chirkin, A. A. Orlov and D. Yu. Paraschuk,
Quantum theory of two-mode interactions in optically anisotropic
media with cubic nonlinearities: generation of quadrature- and
polarization-squeezed light, Quantum Electron. 23, 870 (1993).
\bibitem{Schnabel2003} R. Schnabel, W. P. Bowen, N. Treps, T. C. Ralph,
H.-A. Bachor and P. K. Lam, Stokes operator squeezed continuous
variable polarization states,  Physical Review A 67, 012316 (2003).
\bibitem{Marquardt2007} Ch. Marquardt, J. Heersink, R. Dong, M.V.
Chekhova, A.B. Klimov, L.L. Sanchez-Soto, U.L. Andersen, G. Leuchs,
Quantum reconstruction of an intense polarization squeezed optical
state,quant-ph/0701123, (2007).
\bibitem{Walls1994} D. F. Walls and G. J. Milburn. Quantum Optics, 1.
edition. Springer-Verlag, Berlin, 1994.
\bibitem{Bachor2004}H-A. Bachor and T. C. Ralph. A Guide to Experiments in
Quantum Optics, 2nd edition. Wiley, 2004.
\bibitem{Grangier1987} P. Grangier, R. E. Slusher, B. Yurke and A.
LaPorta, Squeezed light-enhanced polarization interferometer, Phys.
Rev. Lett. 59, 2153 (1987).
\bibitem{Korolkova2002} N. Korolkova, G. Leuchs, R. Loudon, T. C. Ralph
and Ch. Silberhorn, Polarization squeezing and continuous-variable
polarization entanglement, Phys. Rev. A 65, 052306 (2002).
\bibitem{Heersink2003} J. Heersink, T. Gaber, S. Lorenz, O.Glöckl
, N. Korolkova and G. Leuchs, Polarization squeezing of intense
pulses with a fiber-optic Sagnac interferometer, Phys. Rev. A 68,
013815-013827 (2003).
\bibitem{Heersink2005} J. Heersink, V. Josse, G. Leuchs, and U. L.
Andersen Opt. Lett. Vol. 30, No. 10 May 15, 2005
\bibitem{Bowen2002} W. P. Bowen, R. Schnabel, H.-A. Bachor and P. K. Lam,
Polarization Squeezing of Continuous Variable Stokes Parameters,
Phys. Rev. Lett. 88, 093601 (2002).
\bibitem{Josse2003} V. Josse, A. Dantan, L. Vernac, A. Bramati, M. Pinard
and E. Giacobino, Polarization squeezing with cold atoms, Phys. Rev.
Lett. 91, 103601-103604 (2003).
\bibitem{Sherson2006} J. F. Sherson and K. M{\o}lmer, Polarization
squeezing by optical Faraday rotation, Phys. Rev. Lett. 97, 143602
(2006).
\bibitem{Andersen2003}U. L. Andersen and P. Buchhave, Polarization
squeezing and entanglement produced by a frequency doubler, J. Opt.
B. 5, S486S491 (2003).
\bibitem{Andersen2002}  Ulrik L. Andersen and Preben Buchhave, Green bright squeezed light from a cw periodically poled KTP
second harmonic generator, Opt. Express Vol. 10, No. 17 (2002) T.
\bibitem{Tanimura2006} T. Tanimura, D. Akamatsu, Y. Yokoi, A. Furusawa, and M. Kozuma "Generation of a squeezed vacuum resonant on Rubidium D1 line with
periodically-poled KTiOPO$_4$" Opt. Lett. 31, 2344-2346 (2006). G
\bibitem{Hetet2007} G. Hétet, O. Glöckl, K. A. Pilypas, C. C. Harb, B. C. Buchler, H-A. Bachor and
P. K. Lam, Squeezed light for bandwidth-limited atom optics
experiments at the rubidium D1 line, J. Phys. B: At. Mol. Opt. Phys.
40 221-226 (2007)
\bibitem{Takeno2007} Y. Takeno, M. Yukawa, H. Yonezawa and A. Furusawa,
Observation of -9 dB quadrature squeezing with improvement of phase
stability in homodyne measurement, quant-ph/0702139, (2007)
\bibitem{Aoki2006} T. Aoki, G. Takahashi and A. Furusawa
"Squeezing at 946nm with periodically-poled KTiOPO$_4$" Optics
Express 14, 6930-6935 (2006).
\bibitem{BK} G.~D.~Boyd, D.~A.~Kleinman, Parametric interactions of
focused Gaussian light beams, J.~Appl.~Phys., {\bf 39}, 3597 (1968)
\bibitem{PDH} R. Drever, J. Hall, F. Kowalski, J. Hough, G. Ford, A.
Munley and H. Ward, Laser phase and frequency stabilization using an
optical resonator, Applied Physics B: Photophysics and Laser
Chemistry B31 97-105 (1983).
\bibitem{Agarwal2003} G. S. Agarwal and S. Chaturvedi, Scheme to measure
quantum Stokes parameters and their fluctuations and correlations,
J. Mod. Opt. 50, 711 (2003).
\end{thebibliography}

\end{document}